%% file: tvcg_conf_main.tex
\documentclass{vgtc}                          % final (conference style)
\graphicspath{{figures/}{pictures/}{images/}{./}} % where to search for the images

\usepackage{times}                     % we use Times as the main font
\usepackage{tabu}                      % only used for the table example
\usepackage{booktabs}                  % only used for the table example
\usepackage{lipsum}                    % used to generate placeholder text
\usepackage{mwe}                       % used to generate placeholder figures

\usepackage{mathptmx}                  % use matching math font

\input{custom_commands}

\onlineid{ws-AHIMR'26-10}

\vgtccategory{Research}

\vgtcinsertpkg

\title{\PaperTitle{}}

\author{Nuwan Janaka\thanks{e-mail: nuwan.janaka@cityu.edu.hk}\\ %
      \parbox{1.8in}{\scriptsize \centering Synteraction Lab \\ School of Creative Media \& Dept. of Computer Science, City University of Hong Kong \\ Hong Kong, China}
\and Runze Cai\thanks{e-mail: runze.cai@u.nus.edu}\\ %
     \parbox{1.8in}{\scriptsize \centering Synteraction Lab \\ School of Computing, National University of Singapore \\ Singapore} %
\and Yang Chen\thanks{e-mail: cyang@u.nus.edu}\\ %
     \parbox{1.8in}{\scriptsize \centering School of Design, Hong Kong Polytechnic University \\ Hong Kong, China} %
\and Chenyu Zhao\thanks{e-mail: cinderella.zchenyu0912@gmail.com}\\ %
     \parbox{1.8in}{\scriptsize \centering Singapore University of Technology and Design \\ Singapore} %
\and Shengdong Zhao\thanks{e-mail: shengdong.zhao@cityu.edu.hk (Corresponding Author)}\\ %
     \parbox{1.8in}{\scriptsize \centering Synteraction Lab \\ School of Creative Media \& Dept. of Computer Science, City University of Hong Kong \\ Hong Kong, China}}

\abstract{
    \input{paper/0-abstract}
} % end of abstract


\begin{document}

%% The ``\maketitle'' command must be the first command after the
%% ``\begin{document}'' command. It prepares and prints the title block.

%% the only exception to this rule is the \firstsection command
\firstsection{Introduction}

\maketitle

\input{paper/1-introduction}
\input{paper/2-related_work}

\input{paper/3-position}

\input{paper/4-conclusion}

% \section*{Supplemental Materials}
% \label{sec:supplemental_materials}

% Refer to the instructions for this section (\cref{sec:supplement_inst}).
% Below is an example you can follow that includes the actual supplemental material for this template:

% All supplemental materials are available on OSF at \url{https://doi.org/10.17605/OSF.IO/2NBSG}, released under a CC BY 4.0 license.
% In particular, they include (1) Excel files containing the data for and analyses for creating \cref{tab:vis_papers} and \cref{fig:vis_papers}, (2) figure images in multiple formats, and (3) a full version of this paper with all appendices.
% Our other code is intellectual property of a corporation---Starbucks Research---and there is no feasible way to share it publicly.

% \section*{Figure Credits}
% \label{sec:figure_credits}

% Refer to the instructions for this section (\cref{sec:figure_credits_inst}).
% Here are the actual figure credits for this template:

% \Cref{fig:teaser} image credit: Scott Miller / Special to the Vancouver Sun, January 22, 2009, page A6.

% \Cref{fig:vis_papers} is a partial recreation of Fig.\ 1 from \cite{Isenberg:2017:VMC}, which is in the public domain.

%% if specified like this the section will be committed in review mode
\acknowledgments{
GenAI tools have been used to proofread the paper and organize content. The CityU Start-up Grant (9610677) provides partial support.
}

\bibliographystyle{abbrv-doi}

\bibliography{paper/references}

\end{document}

%% file: custom_commands.tex
\newcommand{\PaperTitle}[0]{You Cannot Optimize What You Cannot Measure: Multitasking Evaluation as the Missing Foundation of AI-Mediated Heads-Up Interaction}

\usepackage{tcolorbox}

\newcommand{\TaskCriticality}[0]{\textit{Task Criticality}}
\newcommand{\TaskCoupling}[0]{\textit{Task Coupling}}
\newcommand{\ResourceContention}[0]{\textit{Resource Contention}}

\newcommand{\CriticalityCouplingContention}[0]{$criticality \:\times\:coupling \:\times\: contention$}

\usepackage{subcaption}
\usepackage{multirow}
\usepackage{xcolor}
\usepackage{enumitem}
\usepackage[nomessages]{fp}% http://ctan.org/pkg/fp (math)
\usepackage{soul} % text highlighting
\usepackage{pifont}
\usepackage{longtable}
\usepackage{makecell}
\usepackage{arydshln} % add dash lines
\usepackage{amsmath} 
\usepackage{booktabs}
\usepackage{tabularx} 
\usepackage{subcaption}

\usepackage{transparent}

\usepackage{tabularx}   % Essential for calculating column widths dynamically
\usepackage{xltabular}
\usepackage{ragged2e}   % For better text alignment in narrow columns
\usepackage{array}
\usepackage{longtable}

\newcolumntype{L}{>{\RaggedRight\arraybackslash}X}
\newcolumntype{P}[1]{>{\RaggedRight\arraybackslash}p{#1}}

%% file: paper/0-abstract.tex
% one sentence motivation
% what was done
% what was found (specific)

AI-mediated heads-up augmented reality (AR) replaces fixed interfaces with dynamically adapting ones that decide what information to present, in what form, and when, based on a continually changing context that cannot be fully anticipated beforehand. Although it remains an interface, its behavior over time is only partially specified at design time. We argue that this shift requires a corresponding change in evaluation: from \textit{snapshots} to \textit{trajectories}. A fixed interface is evaluated in a snapshot --- one context, one session, one set of task-performance metrics. A fluid interface must be evaluated over a trajectory --- a sequence of contexts with transitions, sampled from the distribution the interface will actually encounter, and tracked long enough for user trust to form, evolve, and potentially deteriorate.
Drawing on the literature for heads-up AR multitasking enabled by optical see-through head-mounted displays (OST-HMDs), we find that current evaluation practice remains largely snapshot-based. Most studies use fixed-condition, single-session designs; interference between concurrent tasks is rarely quantified directly; and commonly used workload measures cannot disentangle cognitive load attributable to individual tasks. To address these limitations, we argue for three shifts: from isolated metrics to Performance Operating Characteristic (POC) \textit{interference frontiers}, from fixed conditions to evaluation over \textit{context trajectories}, and from single-session snapshots to \textit{longitudinal trust measurement}.

% 'frontier' comes from the concept of a Pareto frontier in economics and human factors engineering.
% The Trade-Off Curve = The Frontier
% If you graph the performance of Task A (e.g., walking speed) on the X-axis and Task B (e.g., reading comprehension) on the Y-axis, you will see a curve representing the absolute best a human can do when balancing both. That outer curve is the frontier.

%% file: paper/1-introduction.tex
% INTRODUCTION (deliver motivation, a bit into second page)
% State of the World
% The big BUT .... (stated as what matters to people)
% Therefore, we did....
% The key findings are...
% The contributions of the work are... (1 to 3)

\label{sec:introduction}

When the interface was a fixed artifact --- a visual layout with pre-authored display rules and a known interaction vocabulary --- a designer's choices could be evaluated once and shipped. The evaluation logic was straightforward: select a context, present the interface to users, and measure task completion time, errors, workload, and usability. One context, one session, and one set of measurements. The interface remained unchanged, and the evaluation captured its performance at that point.

An AI-mediated AR interface differs in one structural way: the designer no longer authors a single, knowable surface. Instead, the interface adapts at runtime --- choosing what to surface, suppress, transform, or delay --- in response to context that shifts from moment to moment: walking or stationary, alone or in conversation, low-load or saturated. It is still an interface. It still has surface-level properties that a user perceives and interacts with. But its behavior across those moments is, by design, not fully known at design time: the designer specifies the adaptation logic, not the set of states it will produce in every possible context. Every adaptation carries an implicit claim --- \textit{this delivery, now, is better for this user than the alternatives} \cite{zhao_heads_up_2023} --- and the designer cannot verify every such claim in advance because the context space is too large to enumerate.

This is the evaluation gap. Fixed interfaces are evaluated in \textbf{snapshots}. Fluid interfaces must be evaluated over \textbf{trajectories}. The difference is not incremental --- add a few more conditions, extend the session by fifteen minutes --- but structural. A snapshot evaluation answers: \textit{does this design work in this context?} A trajectory evaluation must answer: \textit{does this interface, across the range of contexts it will actually encounter, keep the user within an acceptable performance region, adapt well to transitions, and maintain trust over repeated use?}

The mismatch matters because current AR evaluation practice is firmly in snapshot mode. Drawing on a systematic reading of the optical see-through head-mounted display (OST-HMD) multitasking literature, we identify four patterns that are individually unsurprising but together reveal a structural gap (\Cref{sec:related_work}). We then propose a trajectory-based evaluation framework to close that gap (\Cref{sec:position}).

%% file: paper/2-related_work.tex
\section{What The Literature Measures --- And What It Does Not}
\label{sec:related_work}

The OST-HMD multitasking literature is rich in empirical results spanning decades of work. But its evaluation protocols were designed for the fixed-interface paradigm, and four structural patterns emerge from that commitment.

\subsection{Interference Is Not Measured.} 
The dominant instruments are task completion time, aggregate task workload (NASA-TLX), task accuracy, and the System Usability Scale (SUS)~\cite{merino_evaluating_2020, hughes_evaluating_2025}. These measure each task in isolation or capture the undifferentiated overall load. Explicit dual-task constructs --- dual-task decrement, contention overhead, safety-decrement indices --- appear far less frequently. Many studies report only the digital task, whereas those that report both physical and digital tasks can obscure headline gains on one side while real losses occur on the other. The \textit{GlassMessaging} system \cite{janaka_glassmessaging_2023}, for instance, achieves a 40\% improvement in messaging speed while introducing a measurable decrease in accuracy for concurrent physical activity (96.6\% vs. 99.1\%). This pattern is not a flaw in any single study --- it follows structurally from measuring tasks separately rather than measuring interference between them, and it is invisible to any metric that looks at one task at a time. Adjacent fields recognized this long ago: human-factors research on driving standardized detection-response and occlusion protocols \cite{wickens_engineering_2021, eisma_visual_2018, wickens_attentional_1993} precisely because interference between driving and the secondary task is what determines safety. In a trajectory framework, this gap is decisive: without measuring interference at each context point along a trajectory, you cannot plot where the interface sits on the trade-off surface, let alone whether it stays within an acceptable region as contexts shift.

\subsection{Contexts Are Frozen.} 
The typical study compares fixed interface conditions at a fixed context point in a controlled lab setting. This is appropriate for snapshot evaluation: a fixed interface has a single behavior, so measuring it in a single context is a valid test. But the literature's own strongest findings concern transitions and adaptation --- the very things snapshot designs exclude. Users prefer mixed-initiative adaptation over full automation even when automation predicts accurately \cite{lu_exploring_2022, jannat_exploring_2024}, and layout optimizers that reduce effort nonetheless overwrite users' emergent spatial strategies \cite{cheng_semanticadapt_2021}. These results describe what happens \textit{when context changes}, not what happens at a single context point. A trajectory framework would treat these transitions as signals rather than noise: each context shift is a test of how well the interface's adaptation logic navigates the Performance Operating Characteristic (POC) frontier.

\subsection{Time Is Absent.} 
Typical sessions are short, and very few studies extend beyond a single session. In a snapshot framework, this is sufficient --- a fixed interface does not change across sessions, so one session captures its behavior. But a fluid interface's behavior shifts as user and system co-adapt. Habituation, alarm fatigue, and trust recalibration after errors --- the phenomena that determine whether an adaptive interface earns or erodes trust --- operate on timescales invisible to single-session designs~\cite{lu_exploring_2023, stefanidi_real_time_2022, choi_adaptive_2022}. An interface that performs well at minute ten may become aversive by day three; an interface that initially annoys may earn trust as the user calibrates to its adaptation logic. A trajectory framework requires longitudinal designs because trust is a trajectory, not a point in time.

\subsection{Ground Truth Does Not Decompose.} 
Aggregate workload instruments cannot separate the load attributable to the physical task from the load attributable to the digital task. In a snapshot framework, this is acceptable: the interface does not need per-task load to make decisions. In a trajectory framework where the interface adapts to context, it directly constrains what can be sensed and acted upon. To make this concrete: consider a lab study of an adaptive notification system that uses NASA-TLX as its ground-truth workload signal. A participant reports a NASA-TLX score of 65 after a session containing both a walking task and a notification-triage task (e.g., \cite{lu_glanceable_2020, lu_evaluating_2021}). The score is elevated --- but elevated by the physical demand of walking, the mental demand of triage, or both? If the adaptation logic throttles notifications when workload is high, an aggregate score conflating the two tasks will throttle notifications when either task is demanding, including cases where the digital task is the source of load and pausing it is counterproductive. If user-state estimators are trained against labels that cannot express per-task load, they cannot drive an adaptation that depends on per-task load to decide what to surface and when. The sensing pipeline inherits the snapshot deficit and propagates it into runtime decisions.

\textbf{Summary.}
These four gaps share a common root. The field's evaluation apparatus was built for fixed interfaces, where context variation is noise, single sessions are adequate because the artifact does not change, and aggregate workload is sufficient because no downstream system needs per-task decomposition. A fluid interface inverts every term: context variation is signal, behavior shifts across sessions, and per-task load is the input to a moment-by-moment adaptation decision that must be right in context. The apparatus did not follow the artifact.

\begin{table}[htbp]
\centering
\caption{Task meta-properties with illustrative examples.}
\label{tab:task_meta_properties}
\renewcommand{\arraystretch}{1.3}
\small
\begin{tabularx}{\columnwidth}{@{} >{\bfseries\RaggedRight}p{1.8cm} >{\RaggedRight}p{2.0cm} X @{}}
\toprule
Property & Level & Illustrative Examples \\
\midrule
\TaskCriticality{}
  & Low & Glanceable notifications~\cite{lu_glanceable_2020, lu_evaluating_2021}; entertainment~\cite{jain_ubi_touch_2023} \\
  & Medium & Productivity and message access~\cite{janaka_glassmessaging_2023, lu_in_the_wild_2023}; sense-making~\cite{syiem_addressing_2024} \\
  & High & First-responder operations~\cite{zhang_exploring_2024}; anesthesia supervision~\cite{kuge_design_2021}; AR-guided maintenance with safety hazards~\cite{li_effects_2024} \\
\midrule
\TaskCoupling{}
  & Independent Parallel & Audiobook listening while walking~\cite{tan_audioxtend_2024}; ambient notification while working~\cite{lu_evaluating_2021} \\
  & Sequential/Switching & Navigation while walking~\cite{manakhov_gaze_2024}; interleaved assembly steps~\cite{ariansyah_head_2022}; digital interactions during social interaction~\cite{cai_paraglassmenu_2023} \\
  & Tightly Coupled & AR surgery rehearsal~\cite{necker_nested_2024}; real-time robot control~\cite{lee_investigating_2023} \\
\midrule
\ResourceContention{}
  & Low & Stationary use with simple UI~\cite{guarese_augmented_2020} \\
  & Asymmetric & Walking + text reading~\cite{bai_heads_up_2024}; complex diagnostic review while stationary~\cite{kuge_design_2021} \\
  & Saturation & Safety-critical multitask contexts: firefighting~\cite{zhang_exploring_2024}, driving + texting~\cite{he_texting_2015}, high-pressure assembly~\cite{li_effects_2024} \\
\bottomrule
\end{tabularx}
\end{table}

\subsection{A Task Meta-Properties Framework}
\label{sec:meta_properties}

To structure the design space that a trajectory evaluation must cover, we extract three meta-properties from patterns observed across the literature. These properties --- task criticality, task coupling, and resource contention --- capture the design-relevant character of OST-HMD multitasking (digital and physical tasks) configurations and provide the coordinate system for the benchmark cells in~\Cref{sec:position}.

\textbf{Task Criticality} captures the severity of consequences from failures. Low-criticality tasks (glanceable notifications~\cite{lu_glanceable_2020, lu_evaluating_2021}, entertainment~\cite{jain_ubi_touch_2023}) tolerate missed or mistimed deliveries. Medium-criticality tasks (productivity and message access~\cite{janaka_glassmessaging_2023, lu_in_the_wild_2023}, sense-making~\cite{syiem_addressing_2024}) impose real but contained costs. High-criticality tasks (first-responder operations~\cite{zhang_exploring_2024}, anesthesia supervision~\cite{kuge_design_2021}, AR-guided maintenance with safety hazards~\cite{li_effects_2024}) carry safety or mission-critical stakes where errors are unacceptable.

\textbf{Task Coupling} captures the degree to which the physical and digital tasks are interleaved. Independent-parallel tasks (audiobook listening while walking~\cite{tan_audioxtend_2024}, ambient notification while working~\cite{lu_evaluating_2021}) allow the digital task to float freely in time. Sequential/switching tasks (navigation while walking~\cite{manakhov_gaze_2024}, interleaved assembly steps~\cite{ariansyah_head_2022}, digital interactions during social interaction~\cite{cai_paraglassmenu_2023}) require temporal coordination and resumption. Tightly-coupled tasks (AR surgery rehearsal~\cite{necker_nested_2024}, real-time robot control~\cite{lee_investigating_2023}) demand synchronization where the digital and physical tasks share a common object or timeline.

\textbf{Resource Contention} captures how much the two tasks (digital + physical) compete for shared perceptual, cognitive, and motor resources. Low-contention configurations (stationary use with simple UI~\cite{guarese_augmented_2020}) leave ample spare capacity. Asymmetric contention (walking + text reading~\cite{bai_heads_up_2024}, complex diagnostic review while stationary~\cite{kuge_design_2021}) loads one resource pool heavily while leaving others free. Saturation (firefighting~\cite{zhang_exploring_2024}, driving + texting~\cite{he_texting_2015}, high-pressure assembly~\cite{li_effects_2024}) loads multiple resource pools near capacity, leaving no slack for interface overhead.

\Cref{tab:task_meta_properties} summarizes this framework. These three properties jointly define the axes of the benchmark coordinate system used in \Cref{tab:benchmark} (\Cref{sec:position}), ensuring that evaluation cells span the space of contexts a fluid interface will encounter rather than clustering around one convenient design point.

%% file: paper/3-position.tex
\begin{table*}[hptb]
\centering
\caption{Minimal Benchmark Instantiation (First draft). This proposed scaffold serves as a call to the community to standardize dual-task evaluation across distinct context profiles.}
\label{tab:benchmark}
\small
\begin{tabularx}{\textwidth}{@{} >{\raggedright\arraybackslash}p{3.8cm} X X X X @{}}
\toprule
\textbf{Cell (Crit. / Coupling / Contention)} & \textbf{Physical Task \newline (Primary Metrics)} & \textbf{Digital Task \newline (Secondary Metrics)} & \textbf{Scripted Transition} & \textbf{Reported Statistic} \\ 
\midrule

Low / Independent / Low \newline \textit{Notification triage while walking}~\cite{lu_evaluating_2021} & 
Path following (deviation, obstacle-detection rate, gait stability) & 
Message triage (response time, accuracy) & 
Crowd-density increase at fixed waypoint & 
POC frontier of detection rate $\times$ triage throughput; regret vs. oracle timing \\ \addlinespace

Low--Med / Sequential / Physical-Dominant \newline \textit{Guided assembly}~\cite{ariansyah_head_2022} & 
Assembly kit (step-error rate, rework time) & 
Instruction access (access time, navigation actions) & 
Interruption + forced mid-step resumption & 
Resumption lag; POC frontier of assembly error $\times$ access efficiency \\ \addlinespace

High / Tight / Saturated \newline \textit{AR-assisted surgery}~\cite{necker_nested_2024} & 
Procedural simulator (procedural error, hazard-response time) & 
Registered overlay + alerts (alert detection, false-reliance rate) & 
Simulated tracking loss (fail-safe handling scored) & 
Safety-decrement index; alert detection under degradation \\ \addlinespace

High / Independent / Low (off-diagonal) \newline \textit{Ambient hazard monitoring}~\cite{rowen_through_2019} & 
Manual task with injected rare hazards (miss rate, miss latency) & 
Background monitoring (false-alarm annoyance, trust calibration) & 
Hazard during peak manual load & 
Miss rate $\times$ interruption cost; trust trajectory across $\geq 3$ sessions \\ 

\bottomrule
\end{tabularx}
\vspace{1ex}
\par\raggedright\footnotesize
\textbf{Note}: \textit{Regret} is defined against the empirically measured POC frontier for each cell; \textit{false-reliance rate} is the proportion of trials in which the user acts on a stale or incorrect overlay without independent verification. Each cell mandates co-reporting of primary and secondary metrics and per-task decomposed workload; trust-sensitive cells require $\geq 3$ sessions.
\end{table*}

\section{Trajectory-based Evaluation}
\label{sec:position}

We propose a trajectory-based evaluation framework organized around a small set of canonical dual-task scenarios. The goal is to make trajectory evaluation practical and results comparable across studies by providing a shared evaluation scaffold. The benchmark (\Cref{tab:benchmark}) consists of a small set of canonical scenarios, each defined by the same four ingredients and producing the same three outputs. It is not yet a finished instrument but rather a template that the community can adopt, refine, and extend.

\textit{Ingredients (what each scenario specifies).} 
Every scenario is defined by four elements: (1) a \textbf{physical task} with specific primary metrics (e.g., path-following deviation, assembly step-error rate); (2) a \textbf{digital task} with specific secondary metrics (e.g., message triage accuracy, instruction-access time); (3) at least one \textbf{scripted context transition} that occurs mid-session --- a crowd-density increase, an interruption requiring resumption, a tracking loss, a hazard injection --- forcing the interface to adapt under conditions its designer did not pre-script; and (4) a mapping onto the \textbf{\CriticalityCouplingContention{}} coordinate system (\Cref{tab:task_meta_properties}, \Cref{sec:meta_properties}), so that scenarios collectively span the space of contexts a fluid interface will encounter rather than clustering around one convenient design point.

\textit{Outputs (what every study reports).} 
For each scenario, a study reports three quantities. 
(a) A \textbf{POC frontier} --- the empirically measured set of achievable primary--secondary performance pairs, obtained by systematically varying the interface's adaptation parameters (or by comparing multiple interfaces) and plotting the resulting dual-task outcomes as a curve.  
(b) \textbf{Average Regret}, defined as the mean distance between the interface's observed performance and the frontier across all $N$ context points along the trajectory: $\text{Regret} = \frac{1}{N} \sum_{i=1}^N \text{distance}(P_{\text{observed}}(t_i), P_{\text{frontier}}(t_i))$. To ensure scale invariance across disparate task metrics, distance is computed as the Euclidean distance in a normalized $[0, 1]$ performance space, where both primary and secondary metrics are Min-Max scaled such that $1$ represents optimal theoretical performance. A lower average regret indicates the interface consistently stayed closer to the achievable optimum, independent of trajectory length. 
(c) For trust-sensitive cells, a \textbf{trust trajectory} spanning at least three sessions, with deliberate error injections (e.g., a deliberately mistimed notification) whose recovery is measured.

\textit{How this differs from current practice.} 
Today, each study evaluates interfaces in bespoke, often static contexts, yielding snapshot metrics (e.g., a post-hoc cognitive workload scale) that cannot capture how an interface navigates trade-offs during continuous use. While standardizing tasks and metrics is a baseline requirement for fair comparison, the proposed benchmark goes further by standardizing \textit{trajectories of contextual shifts}. Rather than evaluating a static end-state, every interface is evaluated across a parameterized sequence of dynamic changes---such as a scheduled escalation in primary-task attention demands or shifts in visual anchoring constraints. 

Furthermore, testing AI-mediated interfaces in dynamic environments precludes the use of rigidly scripted, pixel-perfect events. To ensure commensurability without sacrificing realism, the benchmark standardizes the \textit{distribution} of unpredictability, such as defining specific error-injection schedules or fixed variances in AI confidence. Consequently, the output is directly comparable over time: ``interface A achieves 8\% average regret across a high-volatility trajectory and 15\% regret during a sustained recovery phase'' explicitly quantifies the process of adaptation. This reveals continuous coverage gaps that a static ``system usability score of 72'' completely obscures, making visible whether an interface only succeeds in a narrow, highly predictable slice of its operating distribution.

\Cref{tab:benchmark} provides an initial four-cell instantiation of this template, including one deliberately off-diagonal cell (high-criticality ambient monitoring, where stakes are highest and evidence thinnest). \Cref{sec:worked_example} walks through a full protocol for one cell. We offer this as a starting point for the community to adopt, challenge, and extend.

The structure of \Cref{tab:benchmark} is designed to operationalize trajectory evaluation in four concrete ways.

\textbf{First, the multi-cell design ensures broad context coverage.} 
No single scenario captures the full range of environments a dynamic interface will face. The four cells span distinct combinations of criticality, coupling, and contention---from low-stakes notification triage to high-stakes surgery. This forces evaluations to sample from a distribution of contexts rather than relying on a single, convenient point estimate. An interface tested in just one cell yields a static snapshot; characterizing it across all four evaluates its adaptability across a diverse trajectory of context classes.

\textbf{Second, paired metrics mandate continuous interference measurement.} 
Each cell requires co-reporting both physical and digital task metrics, plotted together as a POC frontier. This makes the trade-off surface visible. Without paired metrics, it is impossible to distinguish an interface that genuinely improves overall performance from one that merely shifts the cognitive cost from one task to another. For example, \textit{GlassMessaging}'s 40\% messaging-speed gain~\cite{janaka_glassmessaging_2023} looks like a pure win in a standard table of means. On a POC plot, however, the shift toward higher throughput at the cost of lower physical accuracy becomes explicitly clear.

\textbf{Third, parameterized transitions serve as trajectory anchors.} 
Rather than relying on rigidly scripted events, each cell introduces a standardized \textit{distribution} of context shifts---such as escalating crowd density, unexpected interruptions, or tracking loss. These shifts are not experimental noise; they are the core evaluation signal. A fluid interface's true value lies in its adaptation: Does it gracefully degrade in a crowded space? Does it fail-safe when tracking drops? Measuring the interface's response to these parameterized shifts quantifies its resilience and recovery, which a snapshot evaluation entirely misses.

\textbf{Fourth, reported statistics are trajectory-level, not point-level.} 
Outputs are not simplistic ``condition A vs. B'' time-on-task comparisons. Instead, they are structural metrics: a POC frontier, average regret, and longitudinal trust trajectories. Regret, in particular, is a trajectory-native metric. By calculating the \textit{average} distance from the optimal frontier across all context points, it summarizes how effectively the interface navigated trade-offs over time, independent of session length. This provides the mathematical bridge between static snapshots and true dynamic evaluation.

\subsection{A Worked Example: Adaptive Notification Management} \label{sec:worked_example}

To make the protocol concrete, we walk through the first cell of \Cref{tab:benchmark} (Low / Independent / Low: notification triage while walking~\cite{lu_evaluating_2021}) via a full trajectory evaluation of a \textbf{hypothetical} adaptive notification manager, \textsc{NotifAdapt}. This system schedules messages on an OST-HMD based on estimated interruptibility.

\textit{Ingredients.}  
The physical task is path-following through a corridor with embedded obstacles; primary metrics are obstacle-detection rate and path deviation. The digital task is message triage (respond/defer/dismiss); secondary metrics are triage throughput and accuracy. Instead of a rigidly timed event, the parameterized transition introduces a standardized escalation in crowd density midway through the session, shifting the environment from low to high contention. Workload is captured per task via brief probes at segment boundaries.

\textit{Constructing the frontier.}  
The POC frontier is mapped empirically before evaluating the adaptive system. We test four \emph{fixed} delivery policies---immediate delivery, 30-second batching, 120-second batching, and suppress-until-stationary---across both sparse and crowded segments. Each policy yields a (detection rate, throughput) pair; the convex hull of these pairs forms that segment's frontier. Importantly, the frontiers differ by context. In the sparse segment, immediate delivery achieves a hypothetical\footnote{Values are illustrative, consistent with dual-task decrements reported in \cite{bai_heads_up_2024, janaka_glassmessaging_2023}.} (0.97 detection, 3.1 msg/min). In the crowded segment, this same policy degrades to (0.86, 2.9), whereas suppress-until-stationary maintains a 0.96 detection rate but drops throughput to 0.9 msg/min.

\textit{Scoring the adaptive interface.}  
\textsc{NotifAdapt} is then evaluated along the same trajectory. In the sparse segment, it closely mirrors immediate delivery at (0.96, 3.0), resulting in a normalized distance of 0.04 from the frontier. However, it takes ${\sim}40$ seconds to detect the subsequent transition to a crowded environment. During this lag, it continues immediate delivery, yielding a crowded-segment performance of (0.90, 2.6)---a distance of 0.14 from that segment's frontier. The \emph{average regret} across the two segments is therefore 0.09. This decomposition explicitly localizes the performance loss to the transition lag rather than the steady-state policy. A snapshot evaluation misses this completely: tested only in the sparse segment, \textsc{NotifAdapt} appears near-optimal; tested only post-transition, it looks like a poorly tuned static policy. The trajectory evaluation isolates its true weakness: adaptation \emph{latency}.

\textit{Trust trajectory.} 
Sessions repeat over three consecutive days. In session two, the benchmark dictates a deliberately mistimed delivery during an obstacle encounter. We measure trust longitudinally using a brief post-session scale alongside a behavioral proxy: the user's manual override rate. In our hypothetical, overrides spike from 5\% to 21\% following the injected error, then recover to 9\% by the end of session three. This recovery curve is the core trust deliverable. An interface where the override rate fails to recover has broken user trust in a way no single-session SUS score could ever detect.

\subsection{Limitations and Open Questions}

The proposed trajectory-based framework imposes costs that must be acknowledged. Multi-session studies and parameterized transitions increase the logistical burden compared to single-session snapshot designs. A four-cell benchmark with longitudinal trust tracking represents a substantially larger investment per interface than current practice. This raises a practical question: which AR interfaces actually merit trajectory evaluation? We argue the deciding factor is runtime adaptation. A static AR system evaluated under fixed conditions is adequately served by snapshot methods. The trajectory framework is explicitly designed for systems whose behavior across contexts is not fully specifiable at design time---exactly the dynamic systems the community is actively building.

A second open question concerns ecological validity. Parameterized transitions (e.g., escalating crowd density, tracking loss, hazard injections) successfully operationalize adaptation testing, but they remain approximations of naturalistic context shifts. Field-deployed trajectory studies using passive sensing would yield richer, more ecologically valid data, though they introduce significant challenges regarding experimental control and data noise. Ultimately, the research community will need to negotiate the trade-off between controlled, parameterized shifts and uncontrolled field trajectories.

Third, the POC frontier approach assumes primary and secondary task performance can be meaningfully normalized to a common space to calculate average regret (as in \Cref{sec:worked_example}). Disparate tasks naturally utilize different units (e.g., error rate vs. response time, path deviation vs. throughput). Consequently, the choice of normalization technique---whether min-max scaling, z-scores, or domain-specific weighting---can influence the final regret value. Until the community converges on standard normalization conventions, average regret comparisons across fundamentally different task pairs should be interpreted with caution.

Finally, the benchmark cells in \Cref{tab:benchmark} represent an initial draft. While the underlying coordinate system (\CriticalityCouplingContention{}) identifies eight possible cells, we have instantiated four. The remaining cells (e.g., medium-criticality-independent tasks or low-criticality--tightly-coupled tasks) may prove vital for specific application domains and should be populated as researchers gain experience with the framework. We release this scaffold not as a finalized standard, but as a robust starting point for iterative refinement.

%% file: paper/4-conclusion.tex
\section{Conclusion}
\label{sec:conclusion}

The community seeks to formalize the delivery objective for AI-mediated AR interaction. This formalization is blocked not by a shortage of models but by a mismatch of measurement. The interfaces being built are fluid --- adapting at runtime across shifting contexts --- yet the evaluation apparatus inherited from the fixed-interface era measures only at snapshots. Building trajectory-based benchmarks, POC interference frontiers, decomposable per-task ground truth, and longitudinal trust protocols is the necessary foundation. Only then can fluid AR interfaces be evaluated as what they are rather than as what the prior paradigm's tools can see.